\documentclass[fleqn,usenatbib]{mnras}

\usepackage{newtxtext,newtxmath}

\usepackage[T1]{fontenc}
\usepackage{ae,aecompl}

\DeclareRobustCommand{\VAN}[3]{#2}
\let\VANthebibliography\thebibliography
\def\thebibliography{\DeclareRobustCommand{\VAN}[3]{##3}\VANthebibliography}

\usepackage{graphicx}	
\usepackage{amsmath}	

\usepackage{amssymb}	
\usepackage{gensymb}
\usepackage{color}
\usepackage{natbib}
\usepackage{enumitem}
\usepackage{mathtools}
\usepackage{lipsum}
\usepackage{threeparttable}
\usepackage{textcomp,booktabs}
\usepackage{multirow}
\usepackage{subfigure}
\usepackage{verbatim}

\title[MAXI J1348-630]{Detailed analysis on the reflection component for the black hole candidate MAXI J1348-630}

\author[Nan Jia et al.]{
Nan Jia,$^{1,2}$\thanks{E-mail: nanjia@nao.cas.cn}
Xueshan Zhao,$^{1,2}$
Lijun Gou,$^{1,2}$\thanks{E-mail: lgou@nao.cas.cn}
Javier A. Garc\'ia,$^{3,4}$
Zhenxuan Liao,$^{1,2}$
Ye Feng,$^{1,2}$
Yufeng Li,$^{1,2}$
\newauthor
Yuan Wang,$^{1,2}$
Huixian Li,$^{5}$
Jianfeng Wu$^{6}$
\\
$^{1}$Key Laboratory for Computational Astrophysics, National Astronomical Observatories, Chinese Academy of Sciences, 20A Datun Road, Beijing, 100101, China\\
$^{2}$University of Chinese Academy of Sciences, No.19(A) Yuquan Road, Shijingshan District, Beijing,  100049, China\\
$^{3}$Cahill Centre for Astronomy and Astrophysics, California Institute of Technormlogy, Pasadena, CA 91125, USA\\
$^{4}$Dr. Karl Remeis-Observatory and Erlangen Centre for Astroparticle Physics, Sternwartstr. 7, 96049 Bamberg, Germany\\
$^{5}$School of Science, Jimei University, Xiamen 361021, Fujian Province, China\\
$^{6}$Department of Astronomy, Xiamen University, Xiamen, Fujian 361005, China\\
}

\date{Accepted XXX. Received YYY; in original form ZZZ}

\begin{document}
\label{firstpage}
\pagerange{\pageref{firstpage}--\pageref{lastpage}}
\maketitle

\begin{abstract}
The black hole candidate MAXI J1348-630 was discovered on January 26th, 2019, with the Gas Slit Camera (GSC) on-board \textit{MAXI}. 
We report a detailed spectral analysis of this source by using the archived data of \textit{NuSTAR}.
A total of 9 observations covered the complete outburst evolution of MAXI J1348-630 from the hard state to the soft state and finally back to the hard state.  Additionally, the intermediate state is found in the transition from the hard state to the soft state. We use the state-of-art reflection model \verb'relxill' family to fit all the 9 spectra, and the spectra from two focal plane module detectors of \textit{NuSTAR} are jointly fitted for each observation. In particular, we concentrate on the results of the black hole spin parameter and the inclination of the accretion disk. Based on the analysis of the inner radius of the accretion disk, we obtain the spin parameter $a_* =0.78_{-0.04}^{+0.04}$, and the inclination angle of the inner disk $i = 29.2_{-0.5}^{+0.3}$ degrees. Furthermore, we also find that when the black hole is in the hard state, the accretion disk would show a significant truncation. The high iron abundance and ionization of the accretion disk obtained in the fitting results can be possibly explained by the high density of the accretion disk. 
\end{abstract}



\section{Introduction}
\label{sec:intro}

The study of black hole X-ray binary  (BHXRB) is an important research field in astrophysics, and it consists of a black hole and a companion star.
Based on their X-ray light-curve properties, BHXRBs can be classified into two categories: persistent and transient sources. Most are transients, and only a few persistent sources have been found, such as LMC X-1, Cygnus X-1, M33 X-7 and IC 10 X-1. 
MAXI J1348-630 was discovered by the Gas Slit Camera (GSC) on-board \textit{MAXI} \citep{2019ATel12425....1Y} as a new, bright  BHXRB transient at the position RA(J2000) = $13^{\rm{h}}48^{\rm{m}}12^{\rm{s}}.79 \pm 0.03$ and 
Dec(J2000) = -$63^{\rm{d}}16^{\prime}28^{\prime\prime}.48 \pm 0.04$ \citep{2019ATel12434....1K} on January 26th, 2019.
Radio observations of the \textit{Australia Compact Telescope array} (\textit{ATCA}) detected a radio counterpart consistent with the X-ray position \citep{russell2019atca}.
According to the canonical state classification \citep{fender2004towards,homan2005evolution}, MAXI J1348-630 underwent different spectral states, namely, the low/hard state, the hard intermediate state, the soft intermediate state and the high/soft state throughout its outburst.

Since its discovery, several efforts have been made to explore its physical properties. In the previous research, the archived data of \textit{Neutron star Interior Composition Explorer} (\textit{NICER}, \citealt{gendreau2016neutron}) were used to study the outburst evolution and timing properties of MAXI J1348-630 \citep{zhang2020nicer}. Specifically, time lags of the type-B  quasi-periodic oscillations (QPOs) were reported in \citet{belloni2020time} by using \textit{NICER} data of MAXI J1348-630.
A detailed time-lag analysis of MAXI J1348-630 had been carried out by using \textit{Insight}-HXMT over a broad energy band, and they found that  the observed time-lag between the radiations of the accretion disk and the corona leads naturally to the hysteresis effect and the "q"-diagram.\citep{weng2021time}. 
Assuming a face-on disk around a non-spinning black hole, \citet{tominaga2020discovery} fitted spectra obtained by \textit{MAXI}/GSC and \textit{Swift}/XRT, deriving the source distance $D = 4$ kpc and the black hole mass $M = 7$ $M_\odot$.
In another research, using the Australian Square Kilometre Array Pathfinder (ASKAP)
and MeerKAT Data, \citet{chauhan2021measuring} reported H\,{\sc i} absorption spectra of MAXI J1348-630, and determined the source distance to be $2.2_{-0.6}^{+0.5}$ kpc. 
Simultaneous radio (with MeerKAT and \textit{ATCA}) and X-ray (with MAXI and Swift/XRT) observations were used to detect and track the evolution of the compact and transient jets, and the system underwent at least 4 hard-state-only reflares after the main outburst \citep{carotenuto2021black}.
Moreover, the discovery of a giant dust scattering ring around the black hole transient MAXI J1348-630 with \textit{SRG}/eROSITA was reported in \citet{lamer2020giant}, and combining the data from \textit{SRG}/eROSITA, \textit{XMM-Newton}, \textit{MAXI} and \textit{Gaia}, the geometrical distance of MAXI J1348-630 was estimated to be $D = 3.39$ kpc with a statistical uncertainty of 1.1\%. Given the source distance ($D = 3.39$ kpc), the black hole mass was further estimated to be $11\pm2$ $M_\odot$.
In addition, \citet{jana2020accretion}  used TCAF model to fit the combined spectra from \textit{MAXI}/GSC, \textit{Swift}/XRT and \textit{Swift}/BAT, obtaining the black hole mass to be $M = 9.1_{-1.2}^{+1.6}$ $M_\odot$.
\citet{saha2021multi} made a multi-wavelength study of MAXI J1348-630, and they firstly analyzed the broadband spectra of the black hole which covers the radio, optical, ultraviolet and X-ray energy bands.

In a BHXRB, thermal photons emitted from the accretion disk are up-scattered by the hot electrons in the corona, and some of the resulting hard X-ray photons will then be absorbed and reprocessed by the disk, producing the reflection spectrum.
A typical reflection spectrum mainly contains the fluorescent iron K$\alpha$ emission line, absorption edge and Compton hump features \citep{fabian1989x}. 
Under the combined effects of Doppler shifts, beaming and gravitational redshifts close to the black hole, the originally narrow iron K$\alpha$ emission line is broadened to show an asymmetric profile.
The value of the inner edge of the accretion disk can be determined from the profile (especially the red wing) of the broadened iron K$\alpha$ emission line. 
Assuming that the inner radius of the accretion disk would extend to the inner stable circle orbit (ISCO), the spin parameter can be inferred by the relation between spin and ISCO radius \citep{bardeen1972rotating}.
It is generally believed that the inner radius of the accretion disk would extend to the ISCO in the high/soft state \citep{gierlinski2004black,steiner2010constant}.
However, in the hard state, the geometry of the accretion disk is still controversial. Many works have shown that the accretion disk is truncated at a larger radius \citep{gierlinski2008x,tomsick2008broadband,cabanac2009variation,basak2017analysis} and the inner region would be filled by an advection-dominated accretion flow \citep{esin1997advection}, while \citet{rykoff2007swift,reis2010black,xu2018reflection} showed that the accretion disk could possibly extend to ISCO even in the hard state. Therefore, the spin result would be affected when the disk truncation occurs.

Due to the merits of the broad energy band of 3-79 keV,  \textit{NuSTAR} (\citealt{harrison2010nuclear}) is an ideal mission for exploring the black hole reflection properties. In this work, we investigate the archived \textit{NuSTAR} data of MAXI J1348-630 to analyze the black hole reflection component. 
The archived data with 9 spectra cover the complete evolution of the transient outburst, spanning the hard state, intermediate state and soft state.
Using the most sophisticated reflection model to fit the spectra, we can not only constrain the black hole spin parameter and the inclination angle of the inner disk, but also explore the truncation problem of the accretion disk.
Currently, a high-density version of the reflection model, \verb'relxillD' with the maximum density to be $10^{19}$ $\rm{cm}^{-3}$, is also available, providing a proper tool to investigate the high iron abundance effect.
 
This paper is organized as follows.  In section \ref{sec:obs}, we describe the observations and data reduction of MAXI J1348-630. In section \ref{sec:re}, we present the spectral analysis results. In section \ref{sec:dis}, we discuss the disk truncation problem and different reflection models, and also explore the parameter space of the spin parameter and the inclination angle.
In section \ref{sec:con}, we summarize the results and present our conclusion. 

\section{Observations and data reduction}
\label{sec:obs}

\begin{figure}
    \centering
    \includegraphics[width=8cm]{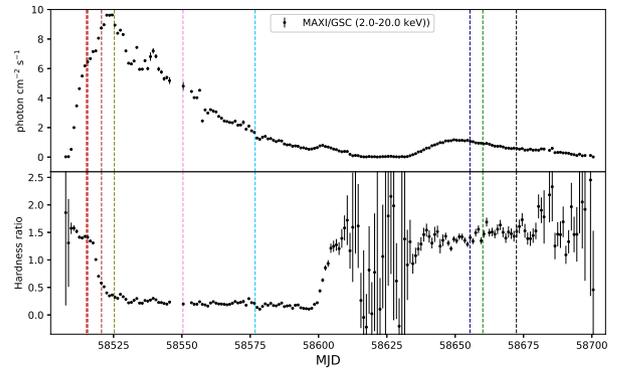}
    \caption{Upper panel: \textit{MAXI}/GSC light curves of MAXI J1348-630 in 2.0-20.0 keV. The colorful dashed vertical line represents the observation of NuSTAR. Lower panel: time evolution of the hardness ratio (4-20 keV/2-4 keV). Since the first two observations are taken very close in time, the two vertical dashed lines on the left side of the figure seems overlapped.}
    \label{fig 1}
\end{figure}

For the first main outburst, the observations of \textit{NuSTAR} started from MJD 58515, and ended in MJD 58627. We obtained the daily averaged light curve from \textit{MAXI}/GSC\footnote{http://maxi.riken. jp} \citep{matsuoka2009maxi}, and the \textit{NuSTAR} observation dates are marked with colorful dashed vertical lines in the light curve in Figure \ref{fig 1}. The hardness ratio plot of MAXI J1348-630 is also shown in the same figure. 
As we can see, there are 9 \textit{NuSTAR} observations in total and, for convenience,  the 9 observations are designated as SP1-SP9, respectively, according to their observation time.
Based on the hardness ratio, we can define the states of the source.
The state of the source, the exposure time and the count rate with different instrument are listed in Table \ref{tab 1}.

We use NuSTARDAS pipeline v2.0.0 with the calibration database (CALDB) version 20210315\footnote{https://heasarc.gsfc.nasa.gov/docs/heasarc/caldb/caldb\_supported\_missions.html} to process the \textit{NuSTAR} data\footnote{https://heasarc.gsfc.nasa.gov/FTP/nustar/data/}. 
The \textit{NuSTAR} source spectra are extracted following the standard procedure\footnote{https://heasarc.gsfc.nasa.gov/docs/nustar/analysis/nustar\_swguide.pdf}, additionally setting \verb'saacalc=2, saamode = OPTIMIZED' and \verb'tentacle = no' to filter background flares caused by enhanced solar activity. We use a circle (with r = $120^{\prime\prime}$) centered on the source to extract the source spectra, and the background spectra are extracted using the same-size circle from a source-free region.
The spectra are then grouped to have at least 25 per bin. In the fitting process, and fitted over the energy band of 3-79 keV.

We use XSPEC v12.11.1\footnote{https://heasarc.gsfc.nasa.gov/xanadu/xspec} to fit the spectra. 
If not specifically mentioned, all uncertainties quoted in this paper are given at 90\% confidence level.

\begin{table}
    \caption{Observation log of MAXI J1348-630}
    \label{tab 1}
    \begin{center}
    \setlength{\tabcolsep}{0.4mm}
    \begin{tabular}{cccccc}
			\toprule
                         \toprule
                         \multicolumn{6}{c}{\emph{\textit{NuSTAR}}}\\
                         \midrule
                         ObsID &  MJD & $\rm{State}^{a}$ & Instrument & Exposure & $\rm{Count}$ $\rm{rate}^{b}$  \\
                         & & & & (ks) & (cts $\rm s^{-1}$)\\
                          \midrule
                        80402315002 & 58515.13 & Hard state & FPMA& 3.04 & 960 \\
                       [1ex]
                            SP1 &  & &  FPMB & 3.21 & 842  \\ 
                         \midrule
                   80402315004      &  58515.47 & Hard state&  FPMA  & 0.48 & 736\\
                  [1ex]
                           SP2 &  &  & FPMB & 3.21 & 778  \\ 
                          \midrule
                    80402315006      &  58520.63 &Intermediate state &  FPMA  & 4.52 & 1278\\
                  [1ex]
                          SP3 &  &  & FPMB & 4.83 & 1182  \\ 
                           \midrule
                    80402315008      &  58525.26 &Soft state &  FPMA  & 4.64 & 1107\\
                  [1ex]
                          SP4 &  &  & FPMB & 5.08 & 1003  \\ 
                            \midrule
                    80402315010      &  58550.37 &Soft state &  FPMA  & 9.71 & 375\\
                  [1ex]
                        SP5  &  &  & FPMB & 10.55 & 331  \\ 
                              \midrule
                      80402315012      &  58576.81 &Soft state &  FPMA  & 12.49 & 122\\
                  [1ex]
                          SP6 &  &  & FPMB & 12.92 & 109  \\ 
                                \midrule
                   80502304002      &  58655.40 & Hard state&  FPMA  & 13.78 & 205\\
                  [1ex]
                        SP7  &  &  & FPMB & 3.21 & 188  \\ 
                                \midrule
                   80502304004      &  58660.05 & Hard state&  FPMA  & 15.37 & 177\\
                  [1ex]
                      SP8  &  &  & FPMB & 15.77 & 163  \\ 
                                \midrule
                   80502304006      &  58672.37 & Hard state&  FPMA  & 17.18 & 117\\
                  [1ex]
                      SP9 &  &  & FPMB & 17.46 & 108  \\

                   \bottomrule
    \end{tabular}
    		\begin{tablenotes}
			\item \textbf{Notes.} $^a$ Classification of spectral states according to \citet{carotenuto2021black}. $^b$ Count rate is measured in 3.0-79.0 keV for FPMA and FPMB respectively.
		\end{tablenotes}
    \end{center}
\end{table}
\section{Spectral analysis and results} 
\label{sec:re}
\begin{figure}
    \centering
    \subfigure[SP1]{
    \includegraphics[width=6cm,angle=270]{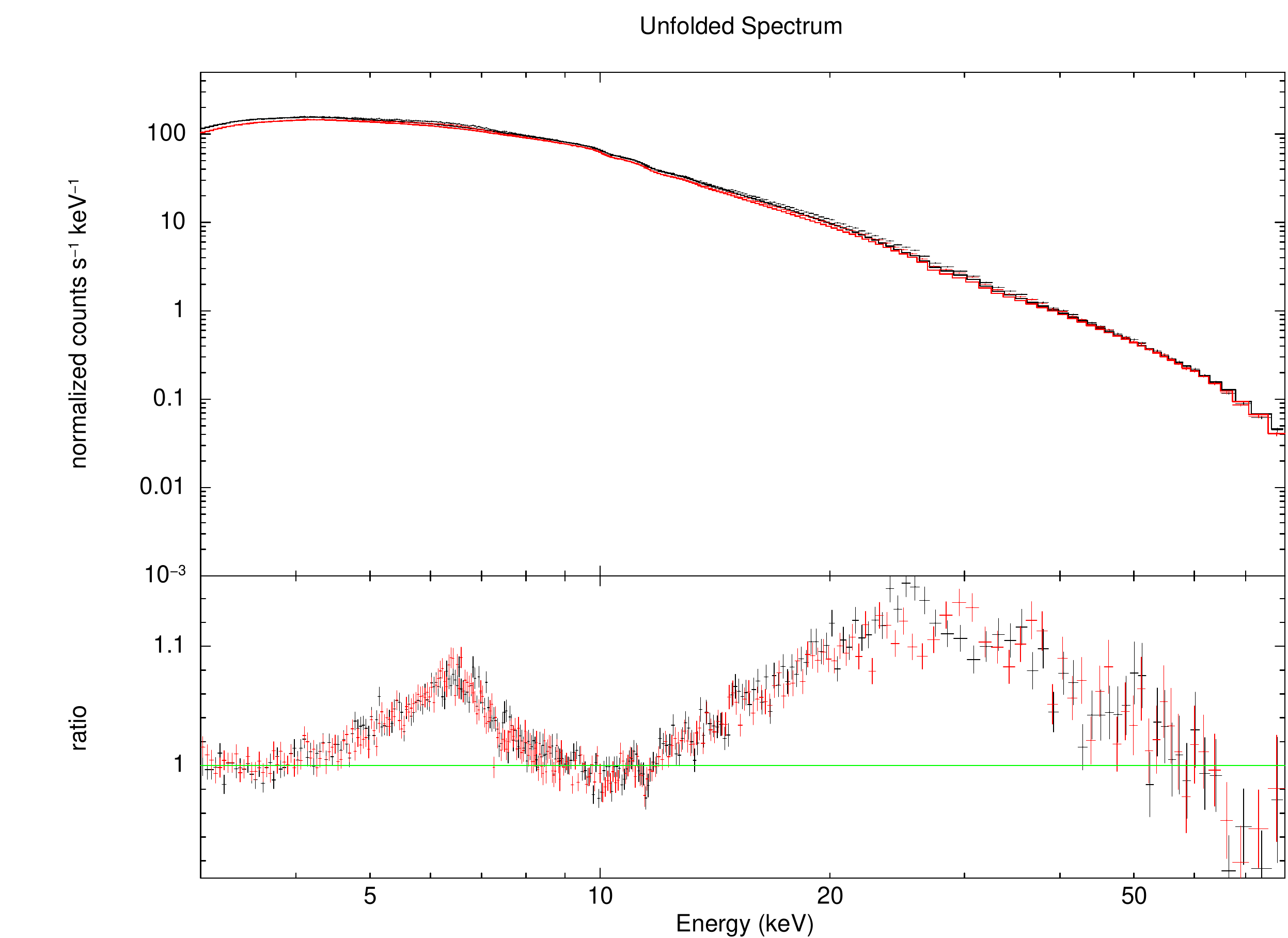}
    }

    \subfigure[SP3]{
    \includegraphics[width=6cm,angle=270]{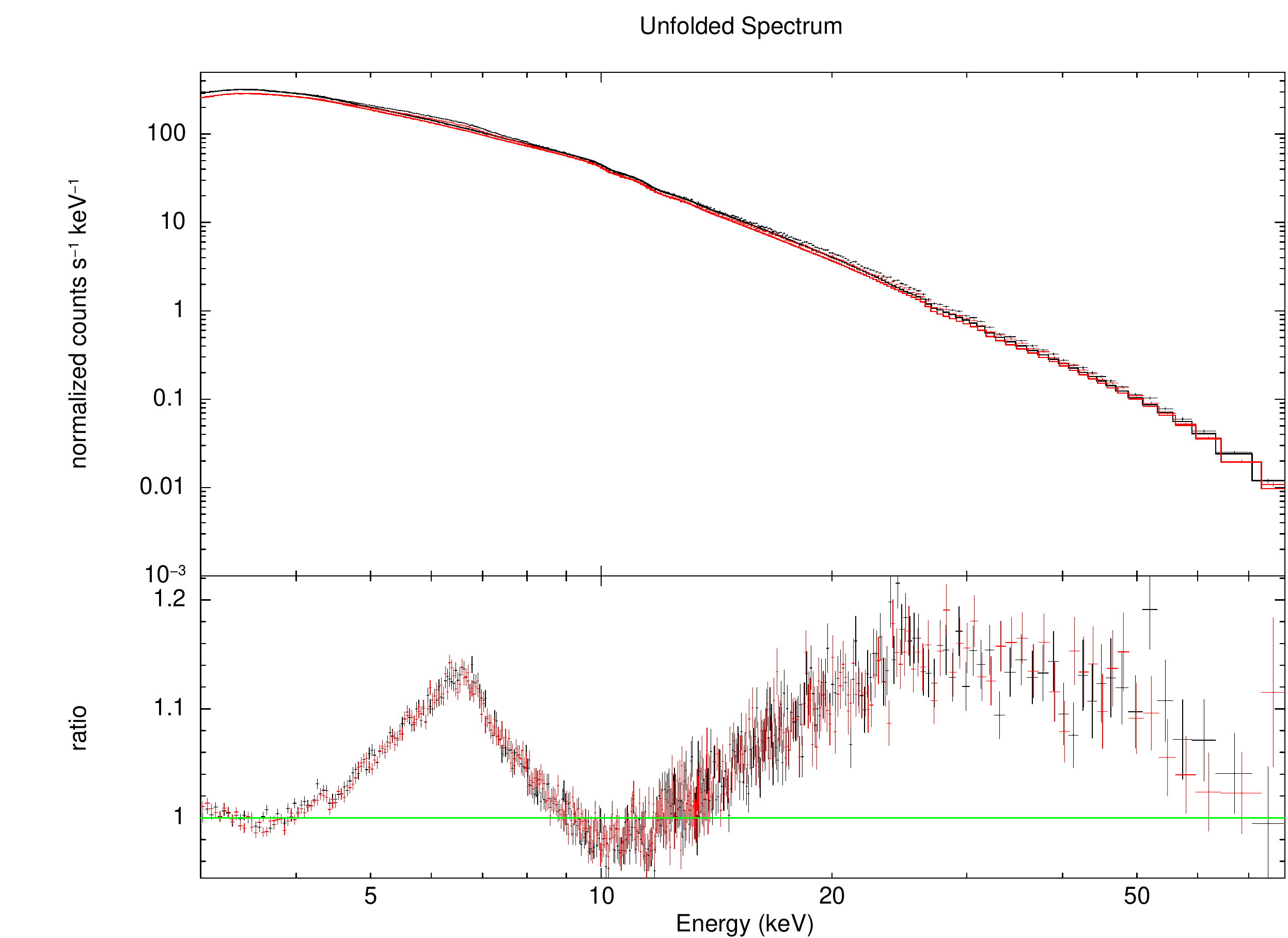}

    }

 \subfigure[SP4]{
    \includegraphics[width=6cm,angle=270]{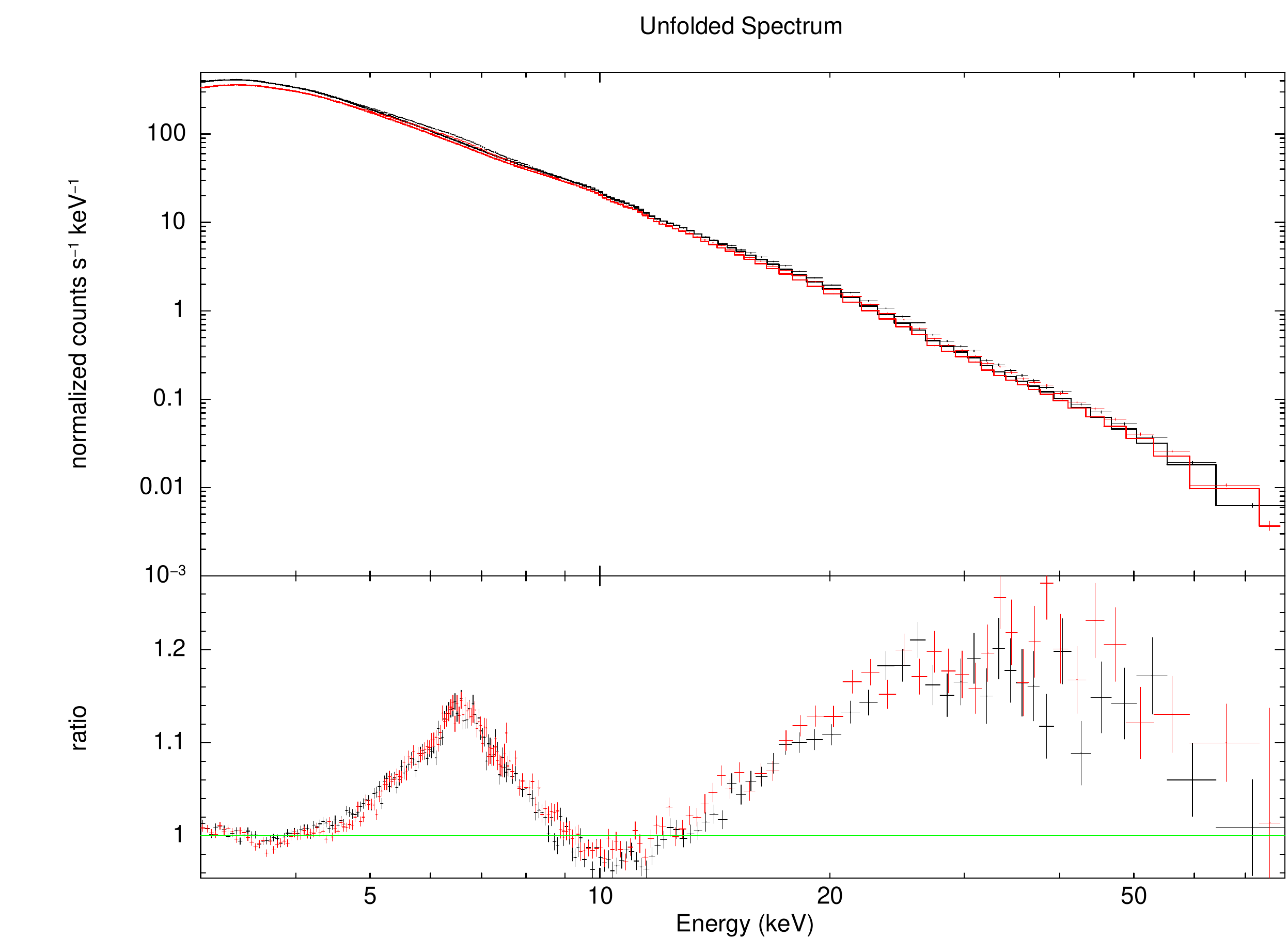}

    }
     \caption{Data/model residuals of the spectra SP1, SP3 and SP4 to the best-fitting result from phenomenological model. FPMA and FPMB data are plotted in black and red, respectively.}
    \label{fig 2}
\end{figure}

For a total number of 9 observations of \textit{NuSTAR}, we use a simple absorbed broken powerlaw model \verb'constant*tbabs*cutoffpl' to fit all the spectra.
The model \verb'constant' is used to reconcile the calibration difference between the FPMA and FPMB detectors, and \verb'tbabs' is the interstellar medium (ISM) absorption model.
We adopt the cross-sections in \citet{verner1996atomic} and abundances in \citet{wilms2000absorption}.
Due to the lack of the spectra below 3 keV, the column density could not be constrained, and we fix the column density at 0.86 $\times$ 10$^{22}$cm$^{-2}$, which is the fitting result using Swift/XRT spectra in \citet{tominaga2020discovery}, When we fit the spectra at this point, we only consider the energy intervals of 3-4, 8-12, 40-79 keV without reflection component effect.
After applying the model, we notice that SP1 and SP2 have an obvious reflection component, which  belongs to the hard state.
SP3 (in the intermediate state) and SP4-6 (in the soft state) present a soft excess feature.
SP7-9 are the observations when the source go back to the hard state.
Considering that the disk thermal radiation would dominate the low energy band, we add the thermal component \verb'diskbb' to the initial model.
It turns out that all the spectra including the hard state spectra (e.g. SP1), the intermediate state spectra (e.g. SP3) and the soft state spectra (e.g. SP4) show a remarkable reflection feature.
From the residual plots, we can clearly see a broad iron line between 5 keV and 7 keV and Compton hump feature above 15 keV. 
The unfolded spectrum and the ratio of data to model are presented in Figure \ref{fig 2}.

Then, we use a preliminary phenomenological model \verb'constant*tbabs*(diskbb+gaussian+cutoffpl)' to fit the spectra.
The model \verb'gaussian'  represents the iron  K$\alpha$ emission line. 
The central energy of the \verb'gaussian' line is frozen at 6.4 keV.
Under the new model configuration, the fitting has been effectively improved,
e.g., SP3 has a reduced $\chi_{\nu}^2$=1.654(= 4346.07/2627).
Although the \verb'diskbb' and \verb'gaussian' could well fit the disk and emission line components, respectively, the Compton hump feature above 15 keV as well as a high-energy cutoff still exist in the data model ratio plot.

Next, we try to replace the phenomenological model \verb'cutoffpl' with the physical model \verb'nthcomp'.
\verb'Nthcomp' is a more realistic description of the continuum shape from the thermal Comptonization than an exponentially cutoff powerlaw.
In this model, the high energy cutoff is sharper than an exponential, and is parameterized by the electron temperature ($k{T}_{\rm{e}}$).
Another major difference is that it incorporates the low energy rollover. 
Using the more reasonable model \verb'nthcomp', the reduced chi-squared value improves slightly.
The fitting statistics is $\chi_{\nu}^2$ =1.618 (= 4252.20/2627).

\begin{figure}
    \centering
    \includegraphics[width=6cm,angle=270]{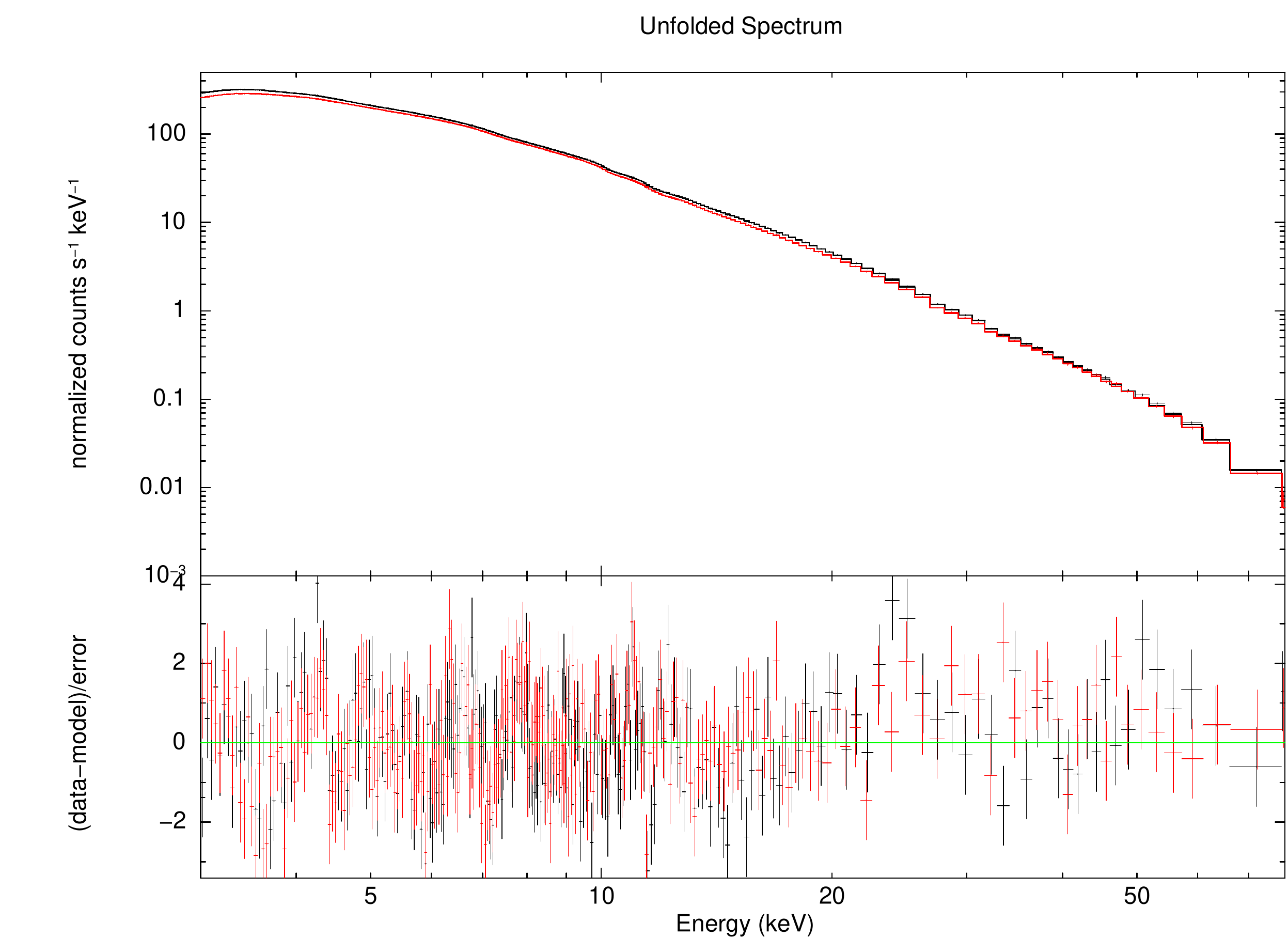}
    \caption{Data/model residuals of the spectrum SP3 to the best-fitting results from model $\tt{const*tbabs*(diskbb+relxillcp)}$. FPMA and FPMB data are plotted in black and red, respectively.}
    \label{fig 3}
\end{figure}

Apparently, although a better fitting result is obtained after using the \verb'nthcomp' model, there still exist some residuals. Therefore we replace the model \verb'nthcomp' with the reflection model \verb'relxillCp'\footnote{http://www.sternwarte.uni-erlangen.de/~dauser/research/relxill} \citep{dauser2014role,garcia2014improved} . 
The complete model is expressed as \verb'constant*tbabs*(diskbb+relxillCp)', which is our final adopted model. 
This sophisticated model \verb'relxillCp' is widely used in black hole binary systems \citep{xu2018reflection,wang2018evolution,you2021insight}. 
It is the combination of the reflection model \verb'xillver'  \citep{garcia2010x,garcia2011x,garcia2013x} and the relativistic model \verb'relline'  \citep{dauser2010broad,dauser2013irradiation}. 
The model of the incident spectrum for the reflection model is \verb'nthcomp', which is a Comptonization continuum. 
The reflection model contains the following parameters: inner radius ($R_{\rm in}$), outer radius ($R_{\rm out}$), break radius ($R_{\rm br}$), which distinguish the inner disk reflection and outer disk reflection; inner index ($q_{\rm in}$), outer index ($q_{\rm out}$), which describe the reflection emissivity in the inner and outer region, respectively; spin parameter ($a_*$), inclination angle ($i$), redshift to the source ($z$), photon index ($\Gamma$), ionization parameter of the accretion disk (in logarithmic scale, ${\rm log}\xi$) that ranges from 0 (neutral) to 4.7 (highly ionized); iron abundance ($A_{\rm Fe}$), electron temperature in the corona ($kT_{\rm e}$), reflection fraction ($R_{\rm f}$) and  normalization (Norm).
Due to the worse constraint on the emissivity index, we assume the canonical case \citep[$q_{\rm out}$ = $q_{\rm in}$ = $3$,][]{fabian1989x}. 
We jointly fit the spectra from FPMA and FPMB detectors for each observation, and the corresponding parameters of the two spectra are linked to each other except the factor constant during each fit. 
We set the outer radius of the accretion disk at its default value $R_{\rm out}$ = $400R_{\rm g}$ (gravitational radius $R_{\rm g}$ = $GM/c^2$).
Since MAXI J1348-630 is a Galactic source, so we set the redshift at $z$ = 0, and all the other parameters are allowed to vary.
After considering the contribution from the reflection component, our adopted final model achieves a much better fit, which can be readily seen from the spectral fit plot as an example for spectrum SP3 in Figure \ref{fig 3}.  The detailed fitting results of all the spectra are listed in Table \ref{tab 2}. From Table \ref{tab 2}, we could see that there exist obvious differences between the fitting results for different states, which we will discuss in detail in the Section \ref{sec:dis}.
Because the fits to the spectra in the intermediate and soft states give the consistent results in the spin parameter (i.e., the inner radius), we thus jointly fit all the spectra in both states to get a better constraint on the spin parameter, and the results are listed in Table \ref{tab 3}.
Assuming $R_{\rm in}$ = $R_{\rm ISCO}$, the spin parameter is measured to be $a_* =0.78_{-0.04}^{+0.04}$, which is our adopted spin parameter for MAXI J1348-630. The inclination angle is also tightly constrained to be $i = 29.2_{-0.5}^{+0.3}$ degrees. 
Our fit shows that the accretion disk is highly-ionized with $\rm{log}\xi$ = $4.33_{-0.04}^{+0.02}$, and the iron abundance is $A_{\rm Fe}>9.05$.

\begin{figure}
    \centering
    \includegraphics[width=8cm]{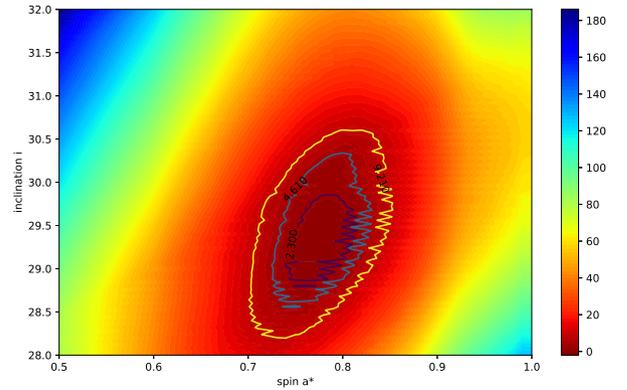}
    \caption{Contours of $\Delta \chi^2$ for the spin parameter $a_*$ and inclination angle $i$. The black, blue, and yellow lines represent the confidence level with $68\%$, $90\%$, and $99\%$ confidence level, respectively.}
    \label{fig 4}
\end{figure}

We also explore the correlation between the spin parameter and the inclination angle with the command  `steppar'. The step size for the spin parameter is set to be 0.012 from 0.4 to 1.0, and  the step size for the inclination angle is 0.1 degrees from 25 degrees to 35 degrees. Then we made a contour plot for the spin parameter and inclination angle (see Figure \ref{fig 4}), and it shows a clear positive correlation between them, which is consistent with the previous works \citep{garcia2018reflection,dong2020spin}.

\begin{table*}
\centering
 \caption{Best-fitting Parameters with Relativistic Models, inner radius is fixed at $R_{\rm{ISCO}}$}
    \label{tab 2}
    \begin{center}
    \setlength{\tabcolsep}{0.3mm}
			\begin{tabular}{ccccccccccccc}
			\toprule
			\toprule
			Spectral & Parameter &SP1&SP2 &SP3 & SP4 
			& SP5& SP6& SP7& SP8& SP9\\
			components & & & & & & \\
			\midrule
			\multicolumn{11}{c}{\emph{\textit{NuSTAR}}}\\
			\midrule
			
			\verb'TBabs' & $N_{\rm{H}}$ $(\times$ 10$^{22}$cm$^{-2}$) 
			&$0.86^\dagger$
			&$0.86^\dagger$
			&$0.86^\dagger$
			&$0.86^\dagger$
			&$0.86^\dagger$
			&$0.86^\dagger$
			&$0.86^\dagger$
			&$0.86^\dagger$
			&$0.86^\dagger$
			
			\\
			\specialrule{0em}{1pt}{1.1pt}
			&&\\
					
		        \verb'Diskbb'
		        &$kT_{\rm{in}}$ (keV)
		        &$0.46_{-0.04}^{+0.05}$
		        &$0.50_{-0.07}^{+0.07}$
		        &$0.68_{-0.01}^{+0.01}$
		        &$0.76_{-0.01}^{+0.01}$
		        &$0.68_{-0.01}^{+0.01}$
		        &$0.57_{-0.01}^{+0.01}$
		        &$0.69_{-0.08}^{+0.07}$
		        &$0.67_{-0.09}^{+0.10}$
		        &$0.61_{-0.16}^{+0.12}$
		        \\
		         [1ex]
		        & 
		        &$1.70_{-0.89}^{+1.98}$
		        &$0.98_{-0.63}^{+2.10}$
		        &$1.94_{-0.11}^{+0.15}$
		        &$2.26_{-0.06}^{+0.04}$
		        &$1.84_{-0.04}^{+0.04}$
		        &$2.35_{-0.06}^{+0.06}$
		        &$0.02_{-0.01}^{+0.02}$
		        &$0.02_{-0.01}^{+0.03}$
		        &$0.02_{-0.01}^{+0.07}$
		        \\	
		        [1ex]
			\verb'relxillCp' 
                   
			& $\textit{a}_{\rm{*}}$ 
			&$0.23_{-0.39}^{+0.26}$
			&$-0.998^{+0.42}$
			&$0.84_{-0.15}^{+0.11}$
			&$0.72_{-0.07}^{+0.09}$
			&$0.60_{-0.21}^{+0.14}$
			&$0.77_{-0.14}^{+0.01}$
			&$-0.70^{+0.92}$
			&$0.09_{-0.70}^{+0.48}$
			&$-0.998^{+1.03}$
			\\
                         [1ex]
			& \textit{i} (deg) 
			&$33.9_{-2.3}^{+2.2}$
			&$27.6_{-4.5}^{+5.0}$
			&$31.7_{-1.6}^{+3.6}$
			&$29.5_{-1.1}^{+1.5}$
			&$27.0_{-1.5}^{+1.4}$
			&$24.8_{-6.3}^{+1.8}$
			&$22.6_{-3.7}^{+7.0}$
			&$28.0_{-4.1}^{+3.8}$
			&$28.1_{-2.2}^{+4.1}$
			 \\
                        [1ex]
			& $\Gamma$
			&$1.694_{-0.008}^{+0.005}$
			&$1.687_{-0.009}^{+0.017}$ 
			&$2.235_{-0.006}^{+0.005}$
			&$2.353_{-0.009}^{+0.007}$
			&$2.195_{-0.005}^{+0.007}$
			&$2.116_{-0.008}^{+0.008}$
			&$1.629_{-0.004}^{+0.012}$
			&$1.623_{-0.007}^{+0.010}$
			&$1.633_{-0.009}^{+0.008}$

			\\
                         [1ex]
			& $\textit{A}_{\rm{Fe}}$ 
                         &$1.96_{-0.29}^{+0.48}$
			&$2.52_{-1.24}^{+1.28}$ 
			&$>8.67$
			&$>7.06$
			&$>9.65$ 
			&$>9.44$
			&$3.75_{-0.84}^{+0.87}$
			&$3.37_{-0.68}^{+0.71}$ 
			&$1.60_{-0.31}^{+0.78}$ 
			\\
                         [1ex]
			& $\rm{log}\xi$ 
                         &$3.30_{-0.05}^{+0.12}$
			&$3.49_{-0.22}^{+0.15}$
			&$>4.63$
			&$4.59_{-0.18}^{+0.05}$
			&$4.32_{-0.04}^{+0.05}$
			&$4.20_{-0.05}^{+0.04}$
			&$3.36_{-0.17}^{+0.10}$
			&$3.21_{-0.07}^{+0.08}$
			&$3.03_{-0.12}^{+0.08}$
			 \\
                         [1ex]
			& $R_{\rm{ref}}$ 
                         &$0.20_{-0.02}^{+0.02}$
			&$0.17_{-0.04}^{+0.04}$
			&$0.54_{-0.05}^{+0.06}$
			&$0.48_{-0.04}^{+0.07}$
			&$0.63_{-0.06}^{+0.06}$
			&$1.10_{-0.13}^{-0.14}$
			&$0.10_{-0.01}^{+0.03}$
			&$0.12_{-0.02}^{+0.03}$
			&$0.15_{-0.02}^{+0.03}$
			 \\
                         [1ex]
			& norm 
                         &$0.194_{-0.005}^{+0.002}$
			&$0.190_{-0.003}^{+0.006}$
			&$0.190_{-0.006}^{+0.006}$
			&$0.125_{-0.004}^{+0.005}$
			&$0.028_{-0.001}^{+0.001}$
			&$0.006_{-0.001}^{+0.001}$
		        &$0.042_{-0.001}^{+0.001}$
			&$0.038_{-0.001}^{+0.006}$
			&$0.028_{-0.002}^{+0.005}$
			 \\
			 [1ex]

			& $k{T}_{\rm{e}}$ 
                         &$25.5_{-1.0}^{+1.2}$
			&$22.5_{-1.6}^{+2.4}$
			&$>225.7$
			&$>140.9$
			&$>296.6$
			&$>205.0$
			&$37.0_{-3.0}^{+5.5}$
			&$22.5_{-1.6}^{+2.4}$
			&$>52.4$
                         \\
                         [1ex]	
                         \hline			
			$C_{\rm FPMB}$ &   &1.02 &1.02& 1.00 & 1.00 & 0.99 & 0.99& 1.00 & 1.01 & 1.01\\
			\hline
			$\chi^2/\nu$ & & 3343.18/3179 &2492.26/239&2944.39/2624 &2518.75/2188 &2340.29/2143  &1815.32/1723 &3523.42/3193 &3381.67/3203  &3282.81/3072\\
			\hline
			$\chi_{\nu}^2$ & & 1.051  & 1.040& 1.122 & 1.151 & 1.092 & 1.054& 1.103 & 1.056 & 1.068 \\ 
			\bottomrule

			\end{tabular}
		\begin{tablenotes}
			\item \textbf{Notes.} The best-fitting parameters obtained by \textit{NuSTAR} observations with model {\tt constant*tBabs*(diskbb+relxillCp)}. The parameters with the symbol “$\dagger$” indicate they are fixed at values given. 
		\end{tablenotes}
		\end{center}
	\end{table*}

\begin{table}
    \caption{Best-fitting Parameters with Relativistic Models, inner radius were frozen at ISCO}
    \label{tab 3}
    \begin{center}
			\begin{tabular}{cccccc}
			\toprule
			\toprule
			Spectral & Parameter & jointly fitted \\
			components & & & & &\\
			\midrule
			\multicolumn{5}{c}{\emph{\textit{NuSTAR}}}\\
			\midrule
			
			\verb'TBabs' & $N_{\rm{H}}$ $(\times$ 10$^{22}$cm$^{-2}$) 
			&$0.86^\dagger$
                       
			\\
			[1ex]
			\verb'Diskbb' 
			& $kT_{\rm{in}}$ (keV) 
			&$0.69_{-0.04}^{+0.04}$(SP3)
			\\
			[1ex]
			&&$0.77_{-0.01}^{+0.01}$(SP4)
			\\
			 [1ex]
			&&$0.66_{-0.01}^{+0.01}$(SP5)
			\\
			 [1ex]
			&&$0.58_{-0.01}^{+0.01}$(SP6)
			\\
                          [1ex]
                          
			&$\rm{N_{Diskbb}}$$(\times 10^4)$
			&$1.70_{-0.04}^{+0.04}$(SP3)
			\\
			[1ex]
			&&$2.23_{-0.03}^{+0.03}$(SP4)
			\\
			 [1ex]
			&&$2.25_{-0.04}^{+0.03}$(SP5)
			\\
			 [1ex]
			&&$2.30_{-0.05}^{+0.04}$(SP6)
			\\
                          [1ex]

			\verb'relxillCp' 

			& $\textit{a}_{\rm{*}}$ 
			&$0.78_{-0.04}^{+0.04}$

			\\
                         \specialrule{0em}{1pt}{1.1pt}
			& \textit{i} (deg) 
			&$29.2_{-0.5}^{+0.3}$

			 \\
                         \specialrule{0em}{1pt}{1.1pt}
			& $\Gamma$ 
			&$2.255_{-0.007}^{+0.005}$(SP3)
			
			\\
			[1ex]
			&&$2.360_{-0.007}^{+0.007}$(SP4)
			\\
			[1ex]
			&&$2.199_{-0.004}^{+0.004}$(SP5)
			\\
			[1ex]
			&&$2.093_{-0.007}^{+0.009}$(SP6)
			\\
                          [1ex]			
                         & $\textit{A}_{\rm{Fe}}$ 
			&$>9.05$ 
			\\
                         [1ex]
			& $\rm{log}\xi$ 
			&$4.33_{-0.04}^{+0.02}$
			 \\
                        [1ex]
			& $R_{\rm{ref}}$ 
			&$0.30_{-0.01}^{+0.01}$(SP3)
			 \\
                         [1ex]
                         &&$0.38_{-0.02}^{+0.01}$(SP4)
			\\
			 [1ex]
			&&$0.76_{-0.04}^{+0.04}$(SP5)
			\\
			 [1ex]
			&&$1.71_{-0.20}^{+0.15}$(SP6)
			\\
                          [1ex]

			& norm 
			&$0.233_{-0.004}^{+0.003}$(SP3)
			 \\
			 [1ex]
			&&$0.139_{-0.003}^{+0.003}$(SP4)
			\\
			 [1ex]
			&&$0.027_{-0.001}^{+0.001}$(SP5)
			\\
			 [1ex]
			&&$0.004_{-0.001}^{+0.001}$(SP6)
			\\
                          [1ex]
 			& $k{T}_{\rm{e}}$ 
			&$193.0_{-82.6}^{+66.8}$(SP3)
                         \\
                         [1ex]
			&&$149.8_{-39.7}^{+94.8}$(SP4)
			 \\
                         [1ex]
			&&$>349.5$(SP5)
			 \\
                         [1ex]
			&&$>270.0$(SP6)
                        \\
                         \\
			\midrule
			
			$C_{\rm FPMB}$ & & 1.00 \\
			\midrule
			$\chi^2/\nu$ & & 10268.64/8693 \\
			\midrule
			$\chi_{\nu}^2$ & & 1.181 \\ 
			\bottomrule

			\end{tabular}
		\begin{tablenotes}
			\item \textbf{Notes.} The best-fit parameters obtained by jointly \textit{NuSTAR} observations for model \verb'const*tbabs*(diskbb+relxillcp)'. The parameters with the symbol “$\dagger$” indicate they are fixed at values given.		
		\end{tablenotes}
		\end{center}
	\end{table}


\section{Discussion}
\label{sec:dis}

\subsection{The truncation of the inner disk}
\label{sec:the}
In section \ref{sec:re}, we use the physical model including \verb'relxillcp' to fit all the 9 X-ray spectra and obtain the relevant parameters of this black hole transient. 
However, among these parameters, we notice that the spin parameter of MAXI J1348-630 varies greatly for different states.
A relatively low spin parameter (corresponding to a larger inner radius of the accretion disk) is obtained for the hard state spectra, while the higher spin parameter (corresponding to a smaller radius of the accretion disk) are derived from the spectra in the intermediate state and soft state.
The standard accretion disk theory holds that the accretion disk will be truncated in the hard state, and would extend to the ISCO radius when it evolves into the soft state and the intermediate state. Clearly our results from fitting the iron line profile follow the trend above, so in the case of MAXI J1348-630,  we believe that the inner disk of the black hole is truncated when it is in the hard state, and the low spin parameter doesn't reflect the actual spin value of the black hole. 

In order to test this picture, we also made another fit by fixing the spin parameter at the maximum value 0.998, and letting the radius of inner disk fit freely in the reflection model. The fitting results are presented in Table \ref{tab 4}, and as shown in Table \ref{tab 4}, the inner disk radius varies from $\sim$ 4 - 10 $R_{\rm ISCO}$ to $\sim$ 2 $R_{\rm ISCO}$ in all cases.  
Apparently the spectra in the hard state give a larger inner radius, and 
the spectra in the intermediate and soft states (e.g. SP3-SP6) gives a relatively smaller inner radius, implying that the disk may have been truncated in the hard state, which is consistent with our the traditional picture for the accretion disk.

Based on the analysis above, we take the spin parameter from the joint fit for the spectra in the intermediate and soft states to be our final spin parameter.

\begin{table*}
\centering
 \caption{Best-fitting Parameters with Relativistic Models, ${a}^*$ is fixed at 0.998}
    \label{tab 4}
    \begin{center}
    \setlength{\tabcolsep}{0.3mm}
			\begin{tabular}{ccccccccccccc}
			\toprule
			\toprule
			Spectral & Parameter &SP1&SP2 &SP3 & SP4 
			& SP5& SP6& SP7& SP8& SP9\\
			components & & & & & & \\
			\midrule
			\multicolumn{11}{c}{\emph{\textit{NuSTAR}}}\\
			\midrule
			
			\verb'TBabs' & $N_{\rm{H}}$ $(\times$ 10$^{22}$cm$^{-2}$) 
			&$0.86^\dagger$
			&$0.86^\dagger$
			&$0.86^\dagger$
			&$0.86^\dagger$
			&$0.86^\dagger$
			&$0.86^\dagger$
			&$0.86^\dagger$
			&$0.86^\dagger$
			&$0.86^\dagger$
			
			\\
			[1ex]
			&&\\
					
		        \verb'Diskbb'
		        &$kT_{\rm{in}}$ (keV)
		        &$0.45_{-0.04}^{+0.04}$
		        &$0.53_{-0.09}^{+0.11}$
		        &$0.67_{-0.01}^{+0.01}$
		        &$0.76_{-0.01}^{+0.01}$
		        &$0.68_{-0.01}^{+0.01}$
		        &$0.57_{-0.01}^{+0.01}$
		        &$0.69_{-0.08}^{+0.07}$
		        &$0.66_{-0.09}^{+0.10}$
		        &$0.53_{-0.11}^{+0.18}$
		        \\
		         [1ex]
		        &$\rm{N_{Diskbb}}$$(\times 10^4)$
		        &$2.04_{-1.06}^{+2.36}$
		        &$0.49_{-0.35}^{+1.80}$
		        &$2.03_{-0.14}^{+0.09}$
		        &$2.27_{-0.06}^{+0.04}$
		        &$1.86_{-0.04}^{+0.04}$
		        &$2.37_{-0.06}^{+0.06}$
		        &$0.02_{-0.01}^{+0.02}$
		        &$0.02_{-0.01}^{+0.03}$
		        &$0.03_{-0.02}^{+0.17}$
		        \\	
		        [1ex]
			\verb'relxillCp' 
                   
			& $\textit{R}_{\rm{in}}$($\textit{R}_{\rm{ISCO}}$) 
			&$3.63_{-0.63}^{+0.90}$
			&$11.91_{-3.27}^{+4.75}$
			&$1.73_{-0.18}^{+0.45}$
			&$2.36_{-0.20}^{+0.19}$
			&$2.63_{-0.21}^{+0.14}$
			&$2.34_{-0.34}^{+0.28}$
			&$6.03_{2.20}^{+1.30}$
			&$4.25_{-1.08}^{+1.70}$
			&$9.70_{-3.80}^{+5.53}$
			\\
                         [1ex]
			& \textit{i} (deg) 
			&$34.5_{-2.2}^{+2.3}$
			&$27.6_{-14.5}^{+4.3}$
			&$32.7_{-0.8}^{+0.8}$
			&$30.6_{-1.1}^{+1.5}$
			&$28.4_{-1.5}^{+1.4}$
			&$25.7_{-1.5}^{+1.4}$
			&$22.9_{-3.4}^{+7.3}$
			&$28.2_{-4.1}^{+3.8}$
			&$27.1_{-3.9}^{+3.4}$
			 \\
                        [1ex]
			& $\Gamma$
			&$1.694_{-0.008}^{+0.005}$
			&$1.703_{-0.035}^{+0.005}$ 
			&$2.236_{-0.005}^{+0.005}$
			&$2.355_{-0.009}^{+0.007}$
			&$2.198_{-0.005}^{+0.007}$
			&$2.117_{-0.008}^{+0.008}$
			&$1.629_{-0.004}^{+0.012}$
			&$1.623_{-0.007}^{+0.010}$
			&$1.641_{-0.017}^{+0.005}$

			\\
                         [1ex]
			& $\textit{A}_{\rm{Fe}}$ 
                         &$2.03_{-0.28}^{+0.47}$
			&$1.14_{-0.21}^{+3.83}$ 
			&$>8.97$
			&$>7.28$
			&$>9.65$ 
			&$>9.38$
			&$3.79_{-0.86}^{+0.87}$
			&$3.39_{-0.65}^{+0.65}$ 
			&$1.01_{-0.13}^{+0.44}$ 
			\\
                         [1ex]
			& $\rm{log}\xi$ 
                         &$3.30_{-0.05}^{+0.12}$
			&$3.30_{-0.03}^{+0.63}$
			&$4.63_{-0.04}^{+0.04}$
			&$4.57_{-0.17}^{+0.05}$
			&$4.30_{-0.03}^{+0.04}$
			&$4.19_{-0.06}^{+0.03}$
			&$3.36_{-0.17}^{+0.10}$
			&$3.21_{-0.07}^{+0.08}$
			&$2.96_{-0.07}^{+0.08}$
			 \\
                         [1ex]
			& $R_{\rm{ref}}$ 
                         &$0.20_{-0.02}^{+0.02}$
			&$0.17_{-0.02}^{+0.04}$
			&$0.57_{-0.10}^{+0.05}$
			&$0.49_{-0.04}^{+0.05}$
			&$0.64_{-0.05}^{+0.05}$
			&$1.12_{-0.15}^{-0.13}$
			&$0.10_{-0.01}^{+0.03}$
			&$0.12_{-0.02}^{+0.03}$
			&$0.17_{-0.01}^{+0.02}$
			 \\
                         [1ex]
			& norm 
                         &$0.194_{-0.004}^{+0.001}$
			&$0.195_{-0.012}^{+0.003}$
			&$0.190_{-0.007}^{+0.008}$
			&$0.127_{-0.005}^{+0.004}$
			&$0.028_{-0.001}^{+0.001}$
			&$0.006_{-0.001}^{+0.001}$
		        &$0.042_{-0.001}^{+0.001}$
			&$0.038_{-0.001}^{+0.001}$
			&$0.033_{-0.007}^{+0.001}$
			 \\
			 [1ex]

			& $k{T}_{\rm{e}}$ 
                         &$25.4_{-1.0}^{+1.3}$
			&$27.6_{-2.9}^{+1.1}$
			&$>227.1$
			&$>131.6$
			&$>350.9$
			&$>270.6$
			&$36.9_{-3.0}^{+5.5}$
			&$42.1_{-5.0}^{+10.2}$
			&$>54.5$
                         \\
                         [1ex]				
			$C_{\rm FPMB}$ &   &1.02 &1.02& 1.00 & 1.00 & 0.99 & 0.99& 1.00 & 1.01 & 1.01\\
			\hline
			$\chi^2/\nu$ & & 3342.52/3179 &2485.22/2395&2945.78/2624 &2509.41/2188 &2343.91/2143  &1810.32/1723 &3522.99/3193 &3381.39/3203  &3282.52/3072\\
			\hline
			$\chi_{\nu}^2$ & & 1.051  & 1.038& 1.123 & 1.147 & 1.094 & 1.050& 1.103 & 1.056 & 1.068 \\ 
			\bottomrule

			\end{tabular}
		\begin{tablenotes}
			\item \textbf{Notes.} The best-fitting parameters obtained by \textit{NuSTAR} observations with model {\tt constant*tBabs*(diskbb+relxillCp)}. The parameters with the symbol “$\dagger$” indicate they are fixed at values given. 
		\end{tablenotes}
		\end{center}
	\end{table*}

\subsection{Reflection flux}
\label{sec:refl}
In this section, we investigate the relative contribution from the reflection component. 
In order to calculate the relative contribution, we have to use the convolution model \verb'cflux' to calculate the flux of each model component.
To calculate the flux of the power law component and the flux of the reflection component separately, we change the model to \verb'const*tbabs*(diskbb+nthcomp+relxillcp)', fixing the $\rm{Refl}\_\rm{frac}$ at -1 in \verb'relxillcp' to represent the reflection component.
In Table \ref{tab 5}, we list the flux of each component as well as the proportion of the reflection component for all the spectra.

\begin{table*}
    \caption{The flux of each model component and reflection component ratio}
    \label{tab 5}
    \begin{center}
    \begin{tabular}{cccccc}
			\toprule
                         \toprule
                         ObsID  &&&Flux (erg/$\rm{cm}^2$/s) & &reflection component ratio \\
                         & \verb'diskbb'  & \verb'nthcomp' & \verb'relxillcp' & total flux  \\
                         \midrule
                       SP1 & 4.349 $\times$ 10$^{-10}$ & 1.011 $\times$ 10$^{-7}$ & 1.479 $\times$ 10$^{-8}$  &1.162 $\times$ 10$^{-7}$ &12.7\%\\
                        [1ex]
                       SP2 & 4.317 $\times$ 10$^{-10}$ & 1.008 $\times$ 10$^{-7}$ & 1.482 $\times$ 10$^{-8}$ &1.162 $\times$ 10$^{-7}$ &12.7\%\\
                       [1ex]
                         SP3 & 8.988 $\times$ 10$^{-9}$ & 4.472 $\times$ 10$^{-8}$ & 2.170 $\times$ 10$^{-8}$ &7.518 $\times$ 10$^{-8}$ &28.8\%\\
                         [1ex]
                         SP4 & 2.364 $\times$ 10$^{-8}$ & 2.230 $\times$ 10$^{-8}$ & 9.850 $\times$ 10$^{-9}$ &5.595 $\times$ 10$^{-8}$ &17.6\%\\
                         [1ex]
                         SP5 & 8.598 $\times$ 10$^{-9}$ & 7.071 $\times$ 10$^{-8}$ & 3.936 $\times$ 10$^{-9}$ &1.955 $\times$ 10$^{-8}$ &20.1\%\\
                         [1ex]
                          SP6 & 3.233 $\times$ 10$^{-9}$ & 1.792 $\times$ 10$^{-9}$ & 1.669 $\times$ 10$^{-9}$ &6.700 $\times$ 10$^{-9}$ &24.9\%\\ 
                           [1ex]
                            SP7 & 1.289 $\times$ 10$^{-10}$ & 2.078 $\times$ 10$^{-8}$ & 1.741 $\times$ 10$^{-9}$ &2.264 $\times$ 10$^{-8}$ &7.7\%\\ 
                             [1ex]
                             SP8 & 9.049 $\times$ 10$^{-11}$ & 1.823 $\times$ 10$^{-8}$ & 1.605 $\times$ 10$^{-9}$&1.992 $\times$ 10$^{-8}$ &8.1\%\\ 
                             [1ex]
                              SP9 & 3.537 $\times$ 10$^{-11}$ & 1.200 $\times$ 10$^{-8}$ & 1.404 $\times$ 10$^{-9}$ &1.346 $\times$ 10$^{-8}$ &10.4\%\\ 
                          \bottomrule
    \end{tabular}

    \end{center}
\end{table*}

As is shown in Table \ref{tab 5}, the reflection component flux ratio of the MAXI J1348-630 in the hard state is only about 10\%, while in the intermedia and soft states, the contribution of the reflection component to the total spectrum can reach up to 20\% even above, which shows that it exists a relatively strong reflection component. Since the reflection feature is quite prominent in the intermediate and soft states, the fitting process is less likely affected by the models, and the corresponding spin parameters should be much more reliable. 

Noted that  usually the reflection component in the high/soft state is relatively weaker compared to the one in the low/hard state, however, the source shows some contrary trend, probably it is related to the exact corona configuration around the black hole, which needs to be investigated in great detail later.

\subsection{High iron abundance}
\label{sec:hia}
In the spectral analysis carried out in Section \ref{sec:re}, we notice that the fitting results in the intermediate and soft states prefer a high iron abundance and a high ionization parameter. 
It is noted that when the high density model of the reflection is applied to some black hole systems, the iron abundance will be lowered \citep{dong2020detailed,chakraborty2021nustar}, indicating the high ionization parameter could possibly be an artifact of the model. Therefore we look into the effect for the source MAXI J1348 as well. 

We replace the reflection model \verb'relxillcp' with \verb'relxillD' in our adopted model above, and
\verb'relxillD'  \citep{garcia2016effects} is the same as the standard reflection model \verb'relxill' but allows a higher density for the accretion disk (between $10^{15}$ to $10^{19}$ $\rm{cm}^{-3}$).
In addition, it has a default value of high energy cutoff at 300 keV, which is consistent with the high electron temperature of the corona in intermediate state and soft state. Except for replacing the reflection model component with a high density version, other components remain the same, so our current model is \verb'const*tbabs*(diskbb+nthcomp+relxillD)'.
We find that after increasing the density of the disk, the iron abundance decrease greatly as well as the ionization has a lower value, as is shown in the Table \ref{tab 6}.  Even though, we notice that all the spin parameters for the high density version of reflection model are consistent with the ones for the default density version. Therefore, our final adopted spin parameter are not affected.

We notice that, compared to the high abundance value in the soft and intermediate states in Table \ref{tab 2}, the abundance value in the hard state is relatively lower, which is probably related to the lower disk component and also the relatively weaker reflection component. As discussed in \citet{garcia2018problem}, the high-density of the accretion disk will contribute to the spectra in both sides: the contributions to the soft-energy component by the free-free heating, and to the line emission and photoelectric absorption. Even though, the large iron abundances required in reflection models to fit the spectra is an open question, and is still under investigation.

\begin{table*}
\caption{Best-fitting Parameters with High Density Relativistic Reflection Models}
\label{tab 6}
\begin{threeparttable}[b]
\begin{center}
			\begin{tabular}{cccccc}
			\toprule
			\toprule
			Spectral & Parameter & SP3 & SP4& SP5& SP6\\
			components & & & & &\\
			\midrule
			\multicolumn{5}{c}{\emph{\textit{NuSTAR}}}\\
			\midrule
			
			\verb'TBabs' & $N_{\rm{H}}$ $(\times$ 10$^{22}$cm$^{-2}$) 
			&$0.86^\dagger$
			&$0.86^\dagger$
			&$0.86^\dagger$
			&$0.86^\dagger$
			\\
                         [1ex]			
                         \verb'Diskbb'
		        &$kT_{\rm{in}}$ (keV)
		        &$0.68_{-0.01}^{+0.01}$
		        &$0.75_{-0.01}^{+0.01}$
		        &$0.67_{-0.01}^{+0.01}$
		        &$0.57_{-0.01}^{+0.01}$
		        \\
		        [1ex]
		        &$\rm{N_{Diskbb}}$$(\times 10^4)$
		        &$1.87_{-0.10}^{+0.03}$
		        &$2.40_{-0.03}^{+0.06}$
		        &$1.90_{-0.04}^{+0.04}$
		        &$2.46_{-0.03}^{+0.06}$
			\\
			[1ex]
			\verb'nthcomp'
			& $k{T}_{\rm{e}}$ 
			&$274.5_{-273.4}$ 
			&$980.9_{-806.8}$
			&$998.4_{-470.4}$
			&$1000.0_{-470.4}$ 
                        \\
                        [1ex]
                        
			& $\Gamma$ 
			&$2.230_{-0.010}^{+0.009}$
			&$2.310_{-0.017}^{+0.010}$
			&$2.189_{-0.009}^{+0.011}$
			&$2.123_{-0.009}^{+0.011}$
			\\
			[1ex]
			 &$\rm{N_{nthcomp}}$
		        &$6.05_{-0.43}^{+0.48}$
		        &$2.48_{-0.43}^{+0.48}$
		        &$0.96_{-0.04}^{+0.03}$
		        &$0.28_{-0.04}^{+0.03}$
			\\
			[1ex]
			\verb'relxillD'

			& $\textit{a}_{\rm{*}}$ 
			&$0.85_{-0.11}^{+0.05}$
			&$0.74_{-0.11}^{+0.05}$
			&$0.64_{-0.16}^{+0.14}$
			&$0.77_{-0.16}^{+0.14}$
			\\
                         [1ex]
			& \textit{i} (deg) 
			&$27.8_{-0.9}^{+1.5}$
			&$18.0_{-4.0}^{+2.4}$
			&$20.6_{-1.6}^{+2.5}$
			&$20.3_{-0.9}^{+1.5}$
			 \\
                           [1ex]
			& $\textit{A}_{\rm{Fe}}$ 
			&$2.34_{-0.45}^{+0.42}$
			&$3.27_{-0.52}^{+2.14}$
			&$5.66_{-0.86}^{+1.84}$
			&$9.05_{-1.41}$ 
			\\
                         [1ex]
			& $\rm{log}\xi$ 
			&$3.70_{-0.03}^{+0.03}$
			&$3.91_{-0.06}^{+0.28}$
			&$3.83_{-0.07}^{+0.13}$
			&$4.01_{-0.07}^{+0.08}$
			 \\
                         [1ex]
			& $R_{\rm{ref}}$ 
			&$-1^f$
			&$-1^f$
			&$-1^f$
			&$-1^f$
			 \\
                         [1ex]
			& $N_{\rm{relxill(Cp)}}$ 
			&$0.128_{-0.014}^{+0.011}$
			&$0.086_{-0.020}^{+0.008}$
			&$0.019_{-0.001}^{+0.001}$
			&$0.007_{-0.002}^{+0.001}$
			 \\
                         [1ex]
			& logN
			&$>18.94$
			&$>18.64$
			&$>18.66$
			&$18.02_{-0.34}^{+0.51}$
                          \\
                         [1ex]

			\midrule
			
			$C_{\rm FPMB}$ & & 1.00 & 1.00 & 1.00 & 1.00 \\
			\midrule
			$\chi^2/\nu$ & & 2925.29/2623  & 2381.33/2187 & 2307.00/2142 & 1800.77/1726\\
			\midrule
			$\chi_{\nu}^2$ & & 1.115 & 1.089& 1.077& 1.043\\
			\bottomrule

			\end{tabular}
		\begin{tablenotes}
			\item \textbf{Notes.} The best-fit parameters obtained by fitting the SP3 spectrum for high density reflection model \verb'const*tbabs*(diskbb+nthcomp+relxillD)'. The parameters with “$\dagger$” indicate they are fixed at values given. 
		\end{tablenotes}
		\end{center}
	\end{threeparttable}
	\label{tab 6}
	\end{table*}

\section{Conclusion}
\label{sec:con}
In this work, we have analyzed the  \textit{NuSTAR} archived data for the black hole MAXI J1348-630, and studied their corresponding spectral properties over the energy range between 3-79 keV.  We find that all the available spectra show a relatively strong reflection feature, therefore,  using our adopted model combination including  a more physical relativistic reflection model \verb'relxillCp', we have fitted all the 9 spectra and obtained a good fit for all the spectra. It shows that we obtain a lower black hole spin value in the hard state, and a relatively higher spin value in the intermediate and soft states, which is consistent with the standard accretion disk theory that the accretion disk will be truncated in the hard state and it will extend to the ISCO radius in the soft state, or occasionally in the intermediate state. Because the spectra in the intermediate and soft states give the consistent results, and we have fitted the spectra in the both states. In this case, the spin parameter is $a_* =0.78_{-0.04}^{+0.04}$ at 90\% statistical confidence, and the inclination angle is $i = 29.2_{-0.5}^{+0.3}$ degrees. 

We also look into the effect of higher accretion disk model to the spin parameter, and it turns out that the spin parameter is not affected, so we take the spin parameter for the standard reflection model above as our adopted one, indicating that MAXI J1348-630 has an intermediate spin.

\section*{acknowledgements}
\label{sec:ack}
This work has made use of data obtained from the \textit{NuSTAR} satellite, a Small Explorer mission led by the California Institute of Technormlogy (Caltech) and managed by NASA's Jet Propulsion Laboratory in Pasadena. 
We thank the \textit{NuSTAR} Operations, Software, and Calibration teams for support with the execution and analysis of these observations.
This research has made use of the \textit{NuSTAR} Data Analysis Software NuSTARDAS, jointly developed by the ASI Science Data Center (ASDC, Italy) and the California Institute of Technormlogy (USA).
L.G. is supported by the National Program on Key Research and Development Project (Grant No. 2016YFA0400804), and by the National Natural Science Foundation of China (Grant No. U1838114), and by the Strategic Priority Research Program of the Chinese Academy of Sciences (Grant No. XDB23040100).
J.W. acknowledges the support of National Natural Science Foundation of China (NSFC grant No. U1938105) and the President Fund of Xiamen University (No. 20720190051). 

\section*{Data Availability}
The data underlying this article are observed by \textit{NuSTAR} which is accessed from 

\noindent \url{https://heasarc.gsfc.nasa.gov/xamin}




\bibliographystyle{mnras}
\bibliography{example} 





\bsp	
\label{lastpage}
\end{document}